\documentclass[%
 a4paper,
 reprint,
superscriptaddress,
nofootinbib,
amsmath,amssymb,
aps,
prb,
]{revtex4-2}

\usepackage{xcolor}
\usepackage{graphicx}% Include figure files
\usepackage{dcolumn}% Align table columns on decimal point
\usepackage{physics}% Necessary for braket notation
\usepackage{verbatim}
\usepackage[hidelinks]{hyperref}% add hypertext capabilities
\usepackage{siunitx}

\begin{document}

% \preprint{APS/123-QED}

\title{Direct and Efficient Detection of Quantum Superposition}% Force line breaks with \\

\author{Daniel Kun}\email[Corresponding author: ]{daniel.kun@univie.ac.at}
\author{Teodor Strömberg}
\author{Michele Spagnolo}
\author{Borivoje Daki\'{c}}
\author{Lee A. Rozema}\email[Corresponding author: ]{lee.rozema@univie.ac.at}\affiliation{University of Vienna, Faculty of Physics, Vienna Center for Quantum Science and Technology (VCQ) {\&} Research Network Quantum Aspects of Space Time (TURIS), Boltzmanngasse 5, 1090 Vienna, Austria}
\author{Philip Walther}\affiliation{University of Vienna, Faculty of Physics, Vienna Center for Quantum Science and Technology (VCQ) {\&} Research Network Quantum Aspects of Space Time (TURIS), Boltzmanngasse 5, 1090 Vienna, Austria}
\affiliation{Christian Doppler Laboratory for Photonic Quantum Computer, Faculty of Physics,  University of Vienna, 1090 Vienna, Austria}

\date{\today}% It is always \today, today,

\begin{abstract}
One of the most striking quantum phenomena is superposition, where one particle simultaneously inhabits different states.
Most methods to verify coherent superposition are indirect, in that they require the distinct states to be recombined.
Here, we adapt an XOR game, in which separated parties measure different parts of a superposed particle, and use it to verify superpositions with \textit{local measurements} and a second independent particle.
We then turn this game into a resource-efficient verification scheme, obtaining a confidence that the particle is superposed which approaches unity exponentially fast.
We demonstrate our scheme using a single photon, obtaining a 99\% confidence that the particle is superposed with only 37 copies.
Our work shows the utility of XOR games to verify quantum resources, allowing us to efficiently detect quantum superposition
without reinterfering the superposed states.
\end{abstract}

\maketitle

\textit{Introduction.}--- The superposition principle, which says that a linear combination of two quantum states is in turn a valid quantum state, lies at the heart of quantum theory and yet challenges our intuition when interpreted literally. Demonstrations of this principle in interference experiments using ever larger quantum states such as molecules~\cite{fein_quantum_2019}, massive nano-particles~\cite{neumeier2022fast}, and even macroscopic states of light~\cite{bruno_displacement_2013}, have continued to attract attention since the advent of quantum theory. 
It is now so routine to place particles, such as photons, in superposition that superposed particles are used as resources for various protocols \cite{del2018two,massa2019experimental,massa2022experimental,BozzioQuantum2020}.
However, in these measurements the particle is never directly observed in a spatial superposition. Thus the description of a superposed quantum state might seem a consequence of the mathematical framework underlying the theory. 
%\danc{We shouldn't phrase it as an "impossibility", since the direct detection of superposition is our main result.}
Already in 1978, Wheeler attempted to address this concern in his famous delayed choice experiment~\cite{wheeler1978past,wheeler1983quantum}.
While strikingly elegant, this thought experiment, as well as its realizations~\cite{hellmuth1987delayed,lawson1996delayed,jacques2007experimental,kim2000delayed} also only constitute indirect demonstrations of the superposition principle, since they rely on interfering the two superposed paths of a test particle with each other. Furthermore, recent works have argued that these experiments admit classical explanations~\cite{catani2023interference}.

\begin{comment}
One could argue that a conclusive validation of the superposition principle was accomplished via the experimental violation of Bell's inequality~\cite{aspect1982experimental,aspect1982experimental2,hensen2015loophole}.
Typical Bell experiments are performed on two particles in a superposition of two different, correlated two-particle states. For example, in the case of photons, a superposition of two polarization states: $\ket{HH}$ and $\ket{VV}$. 
From a quantum mechanical point of view, violating Bell's inequality can only be accomplished if these two-particle states are coherently superposed. 
Bell-type correlations have also been achieved using a {single photon} in a superposition of two spatial modes~\cite{tan1991nonlocality,hardy1994nonlocality,hessmo2004experimental,babichev2004homodyne,fuwa2015experimental,guerreiro_steering_2016}.
%\danc{Lots of detail according to Bori. Sounds like overjustifying.}
Thus, one could conclude that these experiments have validated the superposition principle for single particles.
While this is certainly a valid line of reasoning, it is \textit{indirect}, as it requires an additional assumption that superposed states are required to violate Bell's inequality.
\end{comment}

%Past verifications of the superposition principle have therefore been indirect in one of two ways: they have either required the two parts of the superposition to be reinterfered or required Bell-type violations.
One can ask if the superposition of a \textit{single particle} can be established \textit{directly}, without interfering the two parts of the superposition.
{It could argued that the conclusive validation of the superposition principle was accomplished via the experimental violation of Bell's inequality~\cite{aspect1982experimental,aspect1982experimental2,hensen2015loophole} using a {single photon} in a superposition of two spatial modes ~\cite{tan1991nonlocality,hardy1994nonlocality,hessmo2004experimental,babichev2004homodyne,fuwa2015experimental,guerreiro_steering_2016}. 
However, due to the particle-number superselection rule this cannot be readily extended to massive particles \cite{Heaney2009Bell,Heany2010Bell}.
Moreover, for photons this requires complex homodyne measurements.}
In this Letter, we adapt an XOR game {recently proposed by Del Santo et al.}~\cite{delsanto2020} explicitly designed to detect the presence of coherent superposition.
{In their work,} a classical bound on the probability to win this game is derived by assuming that the information carrier is a classical particle, which is localized in one of two paths.
Exceeding this winning probability thus \textit{directly} reveals the presence of a delocalized (i.e. superposed) particle.
We extend this proposal by measuring a small, fixed number of particles and ask how likely it would be for classical particles to reproduce our observed outcomes. Doing so yields a confidence that the particle is superposed which approaches 1 exponentially fast with the number of measured particles.
{We experimentally demonstrate this protocol by placing} a single photon in a spatial superposition and verify this superposition using only local measurements, and an additional superposed photon.
%In addition to its fundamental interest, our protocol highlights the utility of XOR games for {efficiently} characterizing quantum resources.
\begin{comment}
\begin{figure}[th]
\includegraphics[width=1\linewidth]{fig1-new.pdf}
\caption{\label{fig:setup-cartoon} 
\textbf{Verifying a superposition.}
a) A photon placed in a superposition of two paths using a 50:50 beamsplitter (BS).% can be used as a resource in various quantum protocols.
b) To verify the superposition the two paths are recombined at a second beamsplitter. The interference fringes at the single photon detectors (SPDs) witness superposition.
c) Instead of recombining the two paths, one can use phase-locked local oscillators and photo-diodes (PDs) in each path to violate a Bell inequality, indirectly verifying superposition.
d) Our method makes use of a second superposed ancilla photon.
By interfering the ancilla with the first ``test'' photon and recording coincidence events between detectors, we directly witness superposition.}
\end{figure}
\end{comment}

Our experiment makes use of two single-photon states, independently placed in spatial superpositions. The two parts of the first superposition state are made to interact with the respective parts of the second state on a beamsplitter, forming a \textit{nonlocal interferometer}. Similar to a standard interferometer which encodes phase information in the relative intensity of the two output spatial modes, this nonlocal interferometer encodes the nonlocal phase of the first superposition state in two-photon spatial correlations after the beamsplitters. This allows us to observe this phase without interfering a single photon with itself. To quantify this, we adapt the proposal from Ref.~\cite{delsanto2020} and formulate a two-player XOR game~\cite{brunner_bell_2014} that can be played using our nonlocal interferometer.\footnote{Note that in our work, \textit{nonlocal} refers specifically to the delocalized state formed by the particle placed in superposition, not to be confused with the more familiar notion of Bell nonlocality.}
Such nonlocal interferometers have been proposed \cite{gottesman_longer-baseline_2012, marchese_large_2022} and recently demonstrated \cite{Brown2023Interferometric} to extend the baseline in long-baseline interferometry. 
Similar interferometers have recently also been used to teleport qubits encoded in the Fock basis~\cite{lombardi2002teleportation}, and have been realized on chip \cite{wang2023Deterministic} and with time-bins \cite{Santagiustina2024Experimental}.

%In that context, a faint astronomical object is the source of the superposed photons under test, and its phase encodes information about the astronomical source. A proof-of-principle demonstration of this proposal was recently carried out \cite{Brown2023Interferometric}. A similar interferometer has previously also been used to teleport qubits encoded in the Fock basis~\cite{lombardi2002teleportation}, and been realized on chip with quantum dot single-photons \cite{wang2023Deterministic}.

\begin{figure}[t]
\includegraphics[width=1\linewidth]{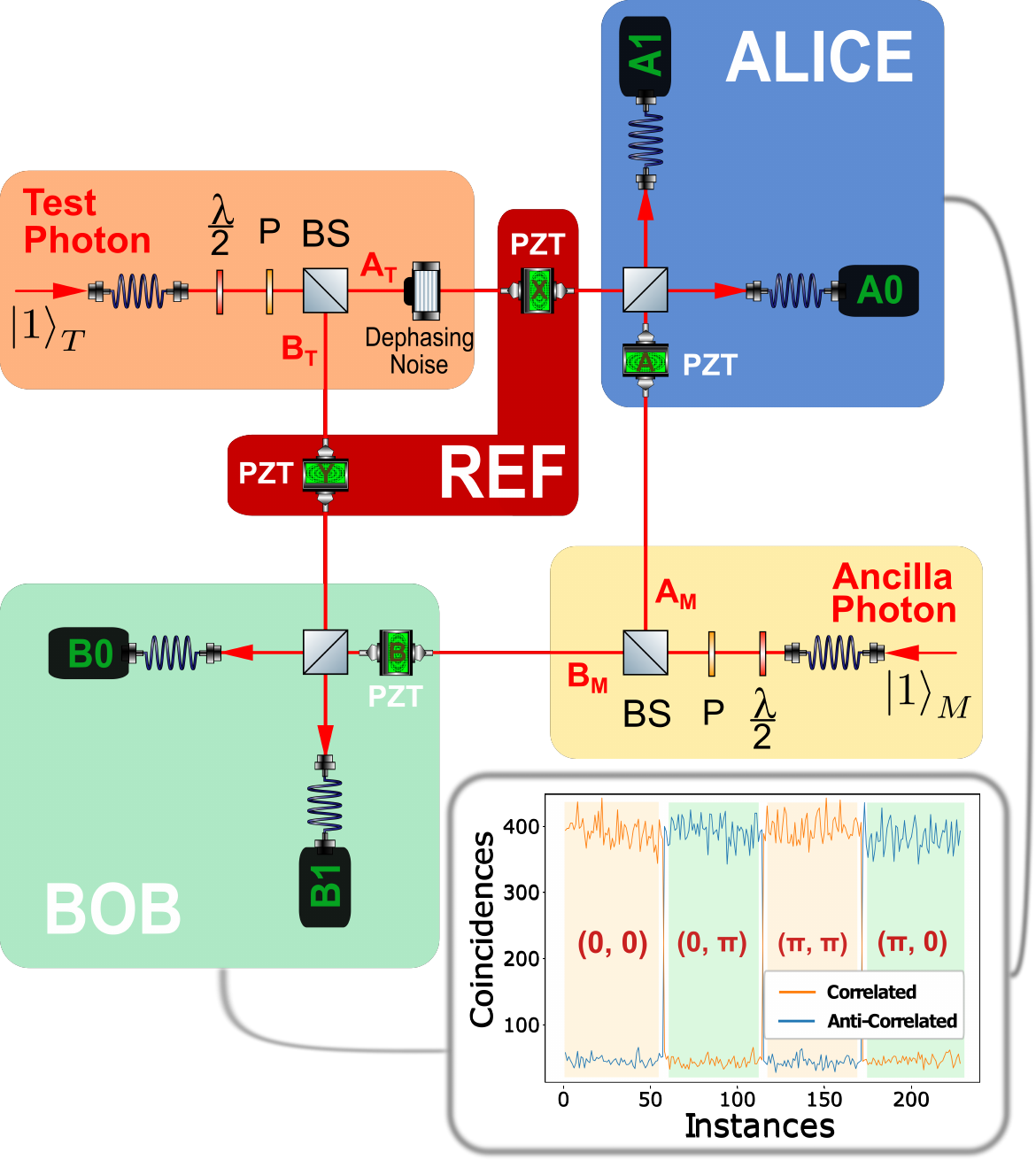}
\caption{
\textbf{XOR Game implementation.} The test (orange field) and ancilla (yellow field) photons are generated via a spontaneous parametric down-conversion photon pair source (not pictured) and then coupled into the nonlocal interferometer, passing through a %\dan{\sout{HWP ($\frac{\lambda}{2}$) and}}
linear polariser (P) to make them indistinguishable in polarization. Each photon is prepared in a coherent superposition of two spatial modes and sent to Alice (blue field) and Bob (green field). The Referee (red field) applies (or not) $\pi$-phases to the spatial modes of the test particle through two piezo-enabled phase delays $X, Y$. Alice and Bob also each control a local phase ($A, B$), which they use to set their shared phase reference. They each locally interfere their test and ancilla modes, recording coincidences between each other's detectors. 
\textbf{Inset: example data run.} 
A plot of correlated (orange) and anti-correlated (blue) detection events as the Referee implements four different phase settings, delimited by shaded regions and indicated in parentheses. Switching the phase setting leads to a switch from correlated to anti-correlated detections. 
{Each x-value corresponds to one ``instance'' of the game, as described in the text.}
}
\label{fig:optical-setup}
\end{figure}

%\LeeC{I think this paragraph could be almost removed/ merged with the protocol section}
Our game consists of a Referee challenging the two players, Alice and Bob, to guess the XOR value of two randomly chosen bits. As illustrated in Fig.~\ref{fig:optical-setup}, the Referee {acts on} a single \textit{test photon} (T) that is sent to Alice or Bob, and encodes the bits by acting on the two respective photon paths. Alice and Bob can locally measure the photon sent by the Referee, and are allowed to exchange classical information. They additionally share a second \textit{ancillary measurement photon} (M), which is in a superposition between their two laboratories and acts as a measurement resource.
In Ref.~\cite{delsanto2020} it was shown that if the test particle is classical, i.e. in a statistical mixture of the two paths, then Alice and Bob can do no better than to randomly guess. 
%\dan{the XOR value}
This is because a classical particle can only contain information about a single bit, since it definitively travels along one of the two paths.
However, if the particle is in a coherent superposition of both paths, Alice and Bob can perform a joint measurement on the test particle and their shared resource state, which allows them to correctly guess the XOR value more often. 

\textit{The protocol.}--- The test photon is prepared in a path superposition state $\ket{\psi_T} \equiv\frac{1}{\sqrt{2}} (\ket{1,0}_{A_T,B_T} + \ket{0,1}_{A_T,B_T})$ by passing it through a 50:50 beamsplitter (see Fig.~\ref{fig:optical-setup}). This is the superposition we wish to verify. The Referee then performs \textit{interventions} on these two paths: $x$ in mode $A_T$ and $y$ in mode $B_T$, where $x,y \in \{0, 1\}$, with 0 (1) denoting the presence (absence) of the intervention.
The Referee now challenges {the players} to produce outputs $a$ and $b$, such that $a \oplus b = x \oplus y$. 
We define Alice's and Bob's \textit{winning probability}
\begin{equation}
    P_{\mathrm{win}} = \sum^1_{\substack{x,y=0 \\ a \oplus b = x \oplus y}} \frac{1}{4}~p(ab|xy),
\end{equation}
for this XOR game, where we have assumed the Referee's choice of interventions $(x, y)$ to be uniformly distributed. 
For classical test particles, Ref.~\cite{delsanto2020} showed that an optimum strategy employed by Alice and Bob will always yield $p(ab|xy) = 1/2 $, and thus $P_{\mathrm{win}} = 1/2$ which corresponds to random guessing. For a quantum superposition, however, Alice and Bob can find a strategy which yields $P_{\mathrm{win}} > 1/2$. 
We will now show, that when the Referee's interventions are $\pi$-phase shifts instead of ``path blockers'' as originally imagined in Ref.~\cite{delsanto2020}, the quantum winning probability goes up to $P_{\mathrm{win}}=3/4$.

To determine the presence of the Referee's interventions, Alice and Bob use an ancillary measurement photon, which is indistinguishable from T.
Similarly to the test photon they prepare it in the superposition state
$\ket{\psi_M} \equiv\frac{1}{\sqrt{2}} (\ket{1,0}_{A_M,B_M} + \ket{0,1}_{A_M,B_M})$.
They then perform joint measurements on the two photons consisting of simple coincidence detections, in contrast to past single-photon Bell violation experiments, which require complex homodyne measurements \cite{hessmo2004experimental,babichev2004homodyne,fuwa2015experimental,guerreiro_steering_2016}.
%\LeeC{Cut the rest of this paragraph:}
%\Lee{We note that a difference between these measurements and the homodyne measurements of Refs.~\cite{fuwa2015experimental,guerreiro_steering_2016} is that their description depends on the interpretation of the input state, while in our scenario the state before the beamsplitters is unambiguously the pure state {$\ket{1}_T\ket{1}_M$}. Any nonlocal correlations observed can therefore only be attributed to the nonlocality of the two individual single-photon states.}

The state of the joint test--ancilla system after the initial beamsplitters is $\ket{\psi_T}\ket{\psi_M}$. As the two photons travel from the beamsplitters to the laboratories of Alice and Bob, the terms corresponding to each spatial mode acquire relative phases. We will denote the phases applied by the Referee by $\varphi_x$ and $\varphi_y$, and set the propagation phases for the ancilla photon to zero for simplicity (see appendix for a discussion). The pre-measurement state is therefore
\begin{equation}
\frac{1}{{\sqrt{2}}} (e^{i\varphi_x}\ket{1,0}_{A_T,B_T} + e^{i\varphi_y}\ket{0,1}_{A_T,B_T})\otimes \ket{\psi_M}.
\end{equation}
To perform their measurements, each party interferes their test and ancilla modes on a 50:50 beamsplitter, and detects which port the photons exit from (blue and green fields in Fig.~\ref{fig:optical-setup}). Alice and Bob then use their detection events to determine the Referee's action. Notice, that half of the time both photons will arrive either in Alice's or Bob's lab. In this case the detection events contain no nonlocal phase information and, again, the best the parties can do is to randomly guess the value of $x \oplus y$. {The other half of the time, both parties receive one photon each. 
In this case, the probability for Alice and Bob's detection events to be correlated or anti-correlated
%(i.e. Alice's detector 0 (1) and Bob's detector 0 (1) click simultaneously)  or an anti-correlated detection event 
%(when Alice's detector 0 (1) and Bob's detector 1 (0) click) 
is complementary and depends on $\varphi_x + \varphi_y$ (see Eqs. A2-A3 in the appendix).
This occurs even though the modes these phases are applied on are not interfered. In other words, the detection events depend nonlocally on these phases.
}

In order to phrase this scenario as an XOR game, we restrict the Referee's phases to  $\varphi_x,\ \varphi_y \in \{0,\pi\}$, and write $\varphi_x = x\pi,\ \varphi_y = y\pi$.
It follows that the anti-correlated events vanish if the Referee's choices satisfy $x\oplus y = 0$. Similarly, if the Referee chooses bits such that $x\oplus y = 1$ the correlated events vanish.
Thus, when Alice and Bob both register a photon, they can win the game by simply outputting the index of the detector that registered a click.
% {However, as previously mentioned, the two photons bunch in one player's laboratory half of the time, and in these cases the best they can do is to randomly guess the Referee's settings.}
In the general case, the probability for Alice and Bob to give outputs $a$ and $b$ given Referee choices $x$ and $y$ is
\begin{equation}
\label{eq:prob-ab-xy}
    p(ab|xy) = \frac{1}{4} \left[ \frac{1}{2} + \cos^2\left(\frac{(x + y)\pi + (-1)^{a \oplus b}\pi}{2}\right)\right].
\end{equation}
Averaging this expression over all settings $x$ and $y$ yields a probability to win the game of $P_{\mathrm{win}}= 3/4$ when Alice and Bob output $a\oplus b$. However, this expression only holds true for perfectly indistinguishable particles in pure quantum states.

\textit{Experimental details.}--- We generate the test and ancilla photons using spontaneous parametric down-conversion (SPDC) in a type-II BBO crystal (see appendix). To implement the interventions, the Referee is given control over two free-space delay stages, which are controlled by piezoelectric transducers (PZTs). Alice and Bob's local measurements are each implemented with a 50:50 beamsplitter and a pair of single-photon detectors ($A_0, A_1$ and $B_0, B_1$ respectively).\footnote{The phase calibration and detector efficiency measurements are described in detail in the appendix.}
Using two photons from an SPDC event ensures high-visibility two-photon interference. Nevertheless, imperfections remain. In the appendix, we compute $P_\mathrm{win}$ in the presence of our main experimental imperfections {for a test particle described by the density matrix}

\begin{equation}
    \rho_T =
    \begin{bmatrix}
   \mathcal{T}_T & \lambda t_Tr_T^* \\
    \lambda t_T^*r_T & \mathcal{R}_T
    \end{bmatrix},
\end{equation}
where $\lambda$ represents the amount of decoherence. The measurement particle state $\rho_M$ is in the analogous pure state with $\lambda=1$. Here $\mathcal{T}_i=|t_i|^2$ and $\mathcal{R}_i=|r_i|^2$ describe the beamsplitters used to superpose the photons $i \in \{T, M\}$.
%The parameter $\lambda$ is related to the purity $\mathcal{P}$ of the superposition state via Eq.~C5.}
The main imperfections are the Hong-Ou-Mandel (HOM) visibility $\mathcal{V}$ between the test and ancilla photons and the imbalance of the two input beamsplitters, which further reduce $P_\mathrm{win}$. The expression for the winning probability accounting for these factors is 
\begin{equation}\label{eq:pwin-lambda-main}
    P_{\mathrm{win}}(\lambda) = \frac{1}{2} + \frac{1}{2}\lambda\mathcal{V}(\mathcal{T}_T\mathcal{R}_M + \mathcal{T}_M\mathcal{R}_T).
\end{equation}
To estimate the expected experimental win rates, we measure the HOM visibility on both detection beamsplitters (shown in Fig.~\ref{fig:hom-dip} in the appendix), finding a visibility of $\mathcal{V} = 94 \pm 2\%$. We also measure the splitting ratio of all beamsplitters, finding that both input beamsplitters have the same $\mathcal{R}:\mathcal{T}$ ratio of $0.65:0.35$, while the detector beamsplitters in Alice's and Bob's labs are balanced within experimental uncertainty. This simplifies Eq.~\ref{eq:pwin-lambda-main} to
\begin{equation}\label{eq:pwin-purity-main}
    P_{\mathrm{win}}(\mathcal{P}) = \frac{1}{2} + \mathcal{V} \sqrt{\frac{\mathcal{RT}}{2}} \sqrt{\mathcal{P}-(\mathcal{R}^2 + \mathcal{T}^2)},
\end{equation}
where we have now replaced $\lambda$ with its expression for purity $\mathcal{P}$ from Eq.~C5.
Setting $\mathcal{P}=1$ gives an expected maximum winning probability of $P_\mathrm{win}=0.7162$.

\begin{figure}[t]
\includegraphics[width=1\linewidth]{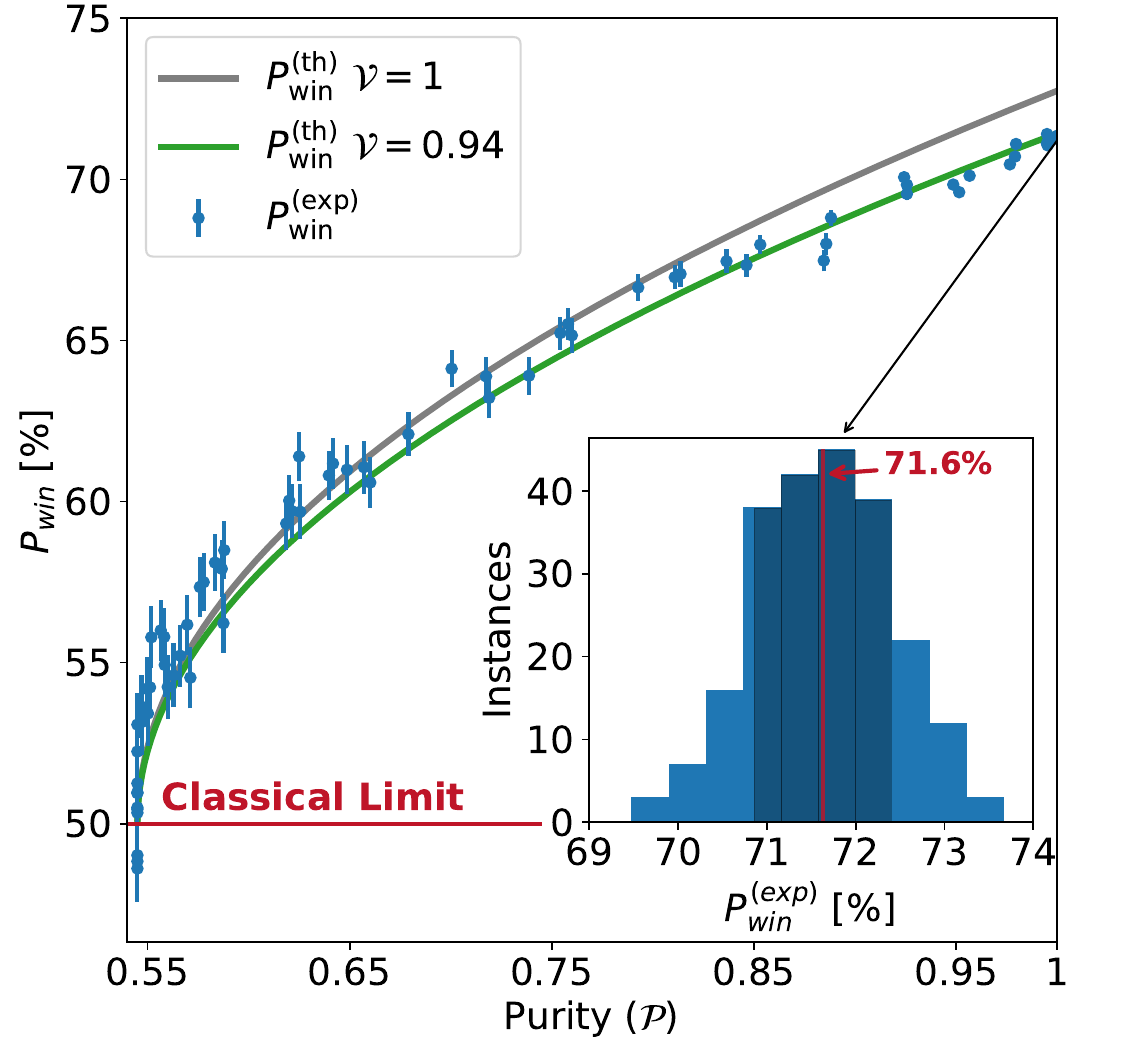}%classical_limit-20230321-10_48_11-to-20230323-05_24_04_pur_with_hist-flipped-transmission.pdf
\caption{ 
\textbf{Probability to win XOR game.}
As the superposition of the test particle is decohered the winning probability $P_{\mathrm{win}}(\mathcal{P})$, approaches the classical limit. Blue dots are experimental data, and the error bars indicate the standard deviation taken over all instances. The average experimental win rate with a pure state is $P^{(\mathrm{exp})}_{\mathrm{win}} = 0.716 \pm 0.007$, well above the classical limit.
The gray curve shows Eq.~\ref{eq:pwin-purity-main} for perfect visibility, while the green curve corresponds to the experimentally measured HOM visibility $\mathcal{V}=94\%$. Good agreement between the model and the data can be seen while the lower bound of 0.54 on the purity is due to imbalance in the preparation beamsplitters.
\textbf{Inset:} {Distribution of experimental} win rates for a single experimental run with a pure state. Each instance contains around 500 played games.}
\label{fig:pwin-pur}
\end{figure}

\begin{figure}[t]
\includegraphics[width=1\linewidth]{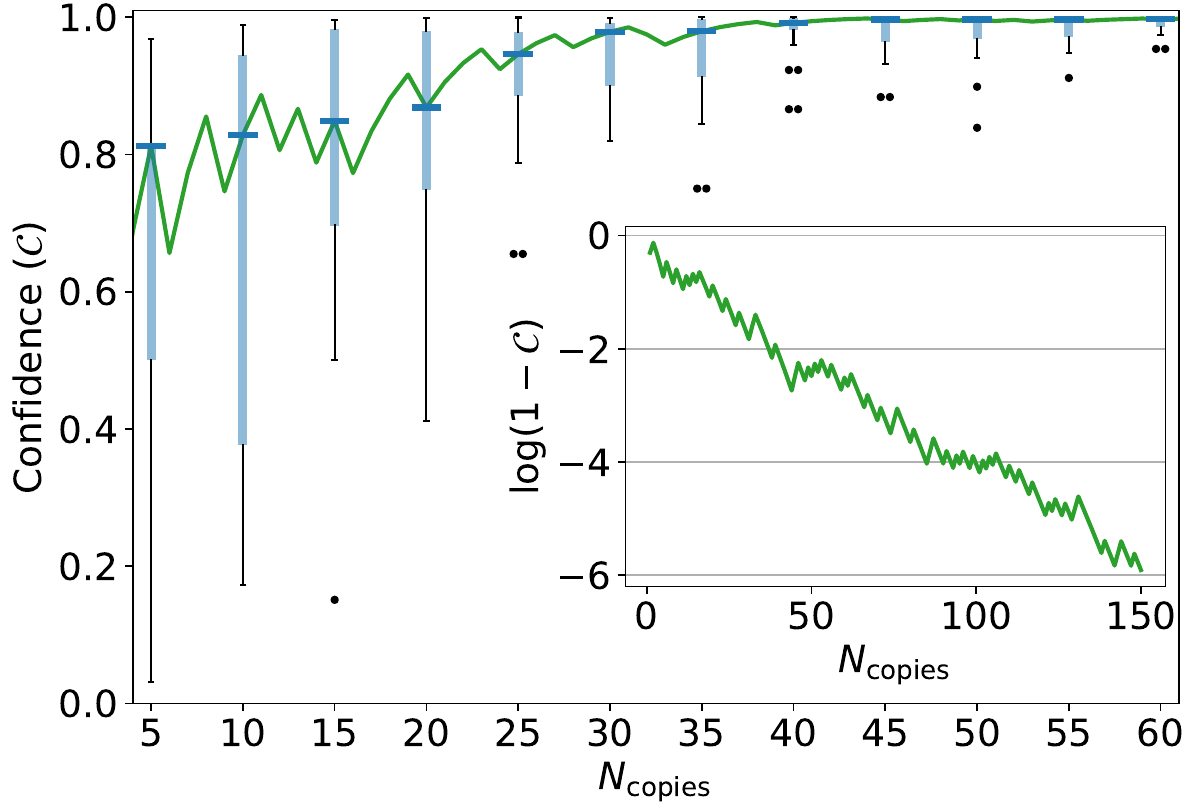}
\caption{\textbf{Efficient confidence estimation.}
{The median confidence, taken over 25 repetitions of the game, versus the number of rounds is indicated by the green line. The box plot illustrates the width of the confidence distributions, and the black dots show the number of outliers at the corresponding confidence value. As shown in the inset, the residual of the median confidence approaches zero exponentially fast in the number of detection events, and 37 copies suffice to reach a confidence above \SI{99}{\percent}.}
%\dan{We play the XOR game with increasing number of copies ($N_\mathrm{copies}$) of the superposition state per experiment and compute the confidence ($\mathcal{C}$). The box plots show the distribution over 25 repeated experimental runs with $N_\mathrm{copies}$. As few as 37 copies already yield a median confidence (red line) of 99\%. The inset shows that the experimental confidence exponentially approaches unity with the number of copies.}
}
\label{fig:conf}
\end{figure}

\textit{Results.}---
Each experimental run consists of 240 instances of the game, with 60 instances for {each phase setting $(\varphi_x, \varphi_y)$.} For each instance, we  acquire coincidence counts for \SI{1}{\second}, yielding approximately 500 coincidence counts per second, distributed across all four coincidence patterns.
Each coincidence count corresponds to one round of the XOR game. One experimental run thus amounts to approximately 120,000 rounds of the XOR game. 
The data in the inset of Fig.~\ref{fig:optical-setup} constitutes one experimental run, where the shaded areas indicate the two XOR sum values. To avoid bias, the order of the four phase settings is determined randomly for each run. 
The analyzed results of one run are displayed in the inset of Fig.~\ref{fig:pwin-pur}.
Therein we see the distribution of the experimental $P_\mathrm{win}$ over the 240 instances. For these data, an average win rate of $0.716 \pm 0.007$ is achieved, which is far above the classical limit of $0.50$, and matches the maximum expected $P_\mathrm{win}$ given by Eq.~\ref{eq:pwin-purity-main}.
Thus we can directly conclude that the test photon is in a coherent superposition.

To study the transition from the quantum regime to the classical limit, we decohere the test photon's spatial superposition by introducing controlled randomness in the test photon phase.
We do so by {adding phase noise with a Gaussian distribution to the Referee's X PZT setting for each instance.}
The standard deviation of the Gaussian distribution, determines the amount of decoherence $\lambda$ and thus the purity of the test photon.
As described in the appendix, we can tune the purity of the test photon {in the range $[0.54,1]$}, where the lower bound is due to the slight imbalance in the spatial superposition state.
We then implement measurement runs, as defined above, for a set of purities in this range.
The resulting win rates are plotted in the main panel of Fig.~\ref{fig:pwin-pur}.
As we vary the purity from $1$ to $0.54$, the win rate decreases according to the predicted experimental $P_{\mathrm{win}}(\mathcal{P})$ from Eq. \ref{eq:pwin-purity-main}.
This prediction, plotted in green in Fig.~\ref{fig:pwin-pur}, agrees well with our experiment, without using any free parameters.
This measurement set further confirms the utility of XOR games for coherence detection, as even low-purity, almost classical superpositions can be effectively verified without the need to reinterfere the spatial modes.

By building on works exploring efficient verification of entanglement~\cite{dimic2018single,saggio2019experimental}, our formulation of the verification task as an XOR game also allows us to verify superpositions efficiently. More concretely, we can express the confidence $\mathcal{C}$ that the test particle is in a superposition as $\mathcal{C} = 1-p$, where 
% \begin{equation}
% \label{eq:pval-main}
    $p= 1 - \sum_{k = 0}^{N_\mathrm{win} - 1}\binom{N}{k} \frac{1}{2^N}$
% \end{equation}
is the p-value for the state not being in a superposition. This p-value corresponds to the probability of a classical particle having generated at least as many wins as experimentally observed (see appendix, Eqs.~E1-E2). The confidence can therefore be interpreted as the probability of the particle having been in a superposition. We evaluate the median experimental confidence over 25 repetitions of the game, and find that in the majority of rounds 37 copies suffice to certify the superposition to \SI{99}{\percent} confidence level (see Fig.~\ref{fig:conf}). 
Moreover, as shown in the inset of the figure, the confidence approaches unity exponentially fast with the number of copies.  

\textit{Discussion.}--- In this work we have demonstrated the superposition principle for a quantum particle using spatially separated local measurements only. 
To do so we created a nonlocal interferometer, wherein the individual phases of a superposed photon are measured nonlocally, without interfering the photon with itself.
%This is a fundamental building block of longer-baseline optical interferometry, which has been proposed as a way to greatly increase the resolution of astronomical optical interferometry~\cite{gottesman_longer-baseline_2012,marchese_large_2022,Brown2023Interferometric}.
Our method to verify superposition can be contrasted with the indirect inference of spatial superposition through single-particle self interference, such as in Young's double slit experiment. {The} experimental apparatus {we employ} is similar to single-photon Bell {tests} or {EPR} steering experiments ~\cite{fuwa2015experimental,guerreiro_steering_2016}, with two crucial differences.
First, the shared resource between the two parties in our work is a delocalized single-photon state, instead of a phase reference set with laser light.
This allows us to stay in the discrete variable picture, and eliminates the need for complex measurements based on homodyne detection.
Second, by designing an XOR game for the task of coherence detection we directly confirm superposition, without having to make the additional assumption that superposition is required to violate a Bell inequality. Finally, the efficiency of our methods is demonstrated by using it to certify quantum superpositions with a confidence that converges to unity exponentially fast.
% \teo{Finally, we demonstrate how our methods can be
% used to efficiently detect the presence of a quantum su-
% perposition, requiring only a limited number of copies of
% the state.}

\begin{acknowledgments}
This research was funded in whole, or in part, by Horizon 2020 and Horizon Europe research and innovation program under grant agreement No 899368 (EPIQUS) and No 101135288 (EPIQUE), the Marie Skłodowska-Curie grant agreement No 956071 (AppQInfo), and the QuantERA II Program under Grant Agreement No 101017733(PhoMemtor). Further funding was received from the Austrian Science Fund (FWF) through 10.55776/COE1 (Quantum Science Austria), 10.55776/F71 (BeyondC) and 10.55776/FG5 (Research Group 5); and from the Austrian Federal Ministry for Digital and Economic Affairs, the National Foundation for Research, Technology and Development and the Christian Doppler Research Association. 
L.A.R. acknowledges support from the Erwin Schrödinger Center for Quantum Science \& Technology (ESQ Discovery).
B.D. also acknowledges support from FWF [P36994-N].
For the purpose of open access, the author has applied a CC BY public copyright license to any Author Accepted Manuscript version arising from this submission.
\end{acknowledgments}

\bibliography{apssamp}% Produces the bibliography via BibTeX.

\onecolumngrid

\appendix

\section{Idealized Interferometer Output State and Phase Reference}

The idealized detection probabilities from Eq.~1 can be derived analytically by applying the unitary transformation of the nonlocal interferometer to the input mode operators.
The input state to the nonlocal interferometer is $\hat{a}_T^\dagger\hat{a}_M^\dagger\ket{0} = \ket{1}_T\ket{1}_M$. The interferometer consists of a pair of beamsplitters, a phase shifter on each arm and another pair of beamsplitters before detection. The resulting output state is then
\begin{align}\label{eq:nlpm-output}
    \ket{\psi^{out}} = 
    &-\frac{\exp(i(\varphi_x + \vartheta_A))}{4}   \underbrace{ \left((\hat{a}^\dagger_0)^2 + (\hat{a}^\dagger_1)^2 \right)}_{\textrm{two photons at A}} \ket{0} ~
    + ~~ \frac{\exp(i(\varphi_x + \vartheta_B)) + \exp(i(\vartheta_A + \varphi_y))}{4} ~ \underbrace{\left(\hat{a}^\dagger_0 \hat{b}^\dagger_0 - \hat{a}^\dagger_1 \hat{b}^\dagger_1\right)}_\textrm{{Correlated}} \ket{0} \nonumber \\
    &- ~\frac{\exp(i(\vartheta_B + \varphi_y))}{4} \underbrace{\left((\hat{b}^\dagger_0)^2 - (\hat{b}^\dagger_1)^2 \right)}_{\textrm{two photons at B}} \ket{0} ~
    + ~~i\frac{\exp(i(\varphi_x + \vartheta_B))) - \exp(i(\vartheta_A + \varphi_y))}{4} ~ \underbrace{\left(\hat{a}^\dagger_0 \hat{b}^\dagger_1 + \hat{a}^\dagger_1 \hat{b}^\dagger_0\right)}_\textrm{{Anti-correlated}} \ket{0},
\end{align}
which includes the phase shifts applied by the Referee ($\varphi_x, \varphi_y$) and the phase shifts ($\vartheta_A, \vartheta_B$) applied by Alice and Bob to set a shared phase reference. A more explicit derivation including the unitaries can be found in Section~\ref{sec:pwin-derivation}, where experimental imperfections are modeled.

\begin{figure}[h]
    \centering
    \includegraphics[width=0.5\linewidth]{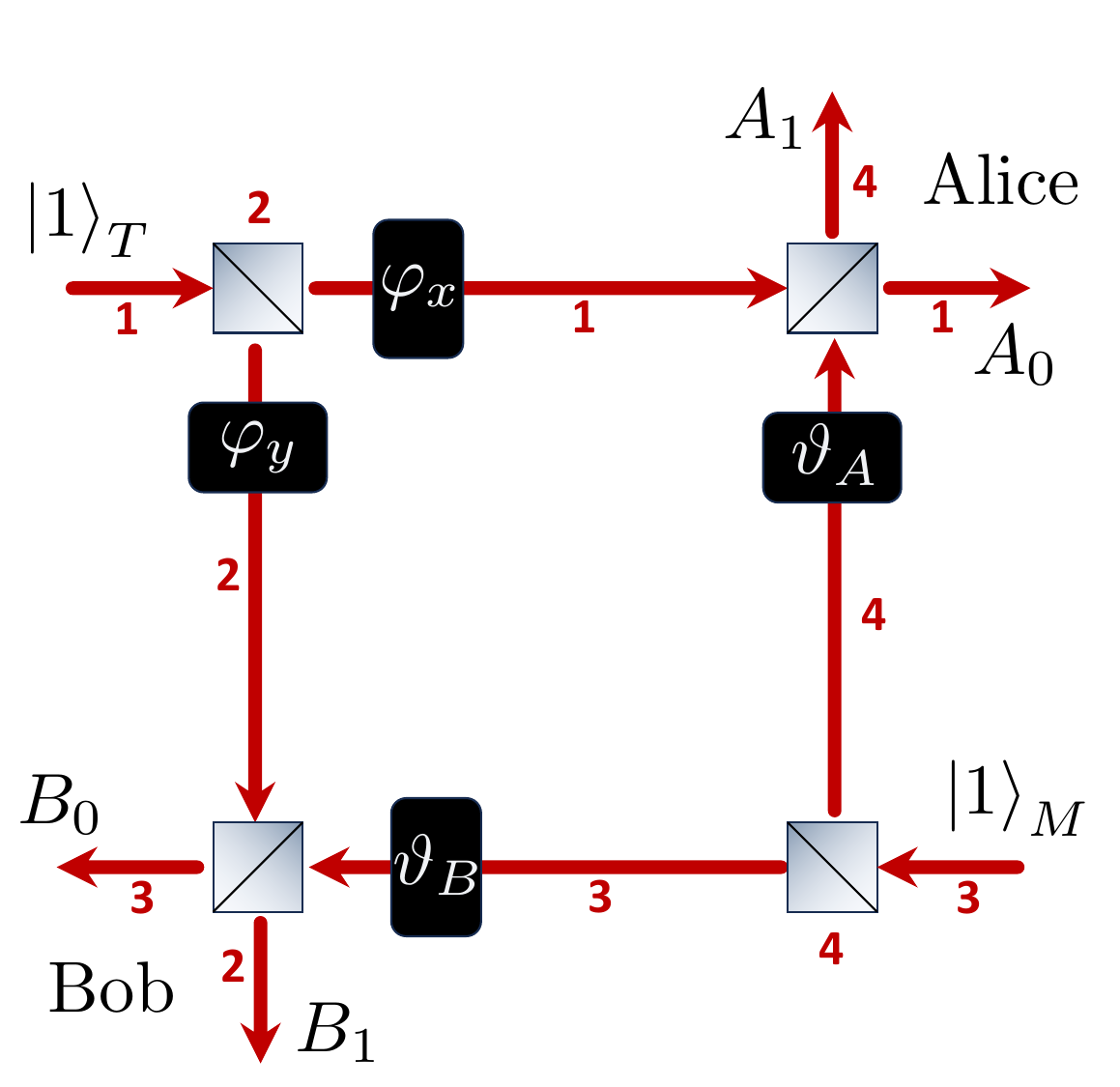}
    \caption{Simplified schematic of setup with labeled modes}
    \label{fig:mode-convention}
\end{figure}

Using Eq.~\ref{eq:nlpm-output} to compute the probabilities for Alice and Bob's detectors to click simultaneously, we find
\begin{align}
    p_{00} &= |\bra{0} \hat{a}_0\hat{b}_0\ket{\psi^{out}}|^2 = \frac{1}{4}\cos^2\left(\frac{\varphi_x + \varphi_y + \vartheta_A + \vartheta_B}{2} \right)	 = |\bra{0} \hat{a}_1\hat{b}_1\ket{\psi^{out}}|^2 = p_{11} \label{eq:corr}\\
    p_{10} &= |\bra{0} \hat{a}_1\hat{b}_0\ket{\psi^{out}}|^2 = \frac{1}{4}\sin^2\left(\frac{\varphi_x + \varphi_y + \vartheta_A + \vartheta_B}{2} \right) = |\bra{0} \hat{a}_0\hat{b}_1\ket{\psi^{out}}|^2 = p_{01}, \label{eq:acorr}
\end{align}
which shows that the probability for the correlated ($p_{00}, p_{11}$) and anti-correlated ($p_{01}, p_{10}$) outputs have a complementary dependency on the sum of all four applied phase terms.

\begin{figure}
    \centering
    \includegraphics[width=1\linewidth]{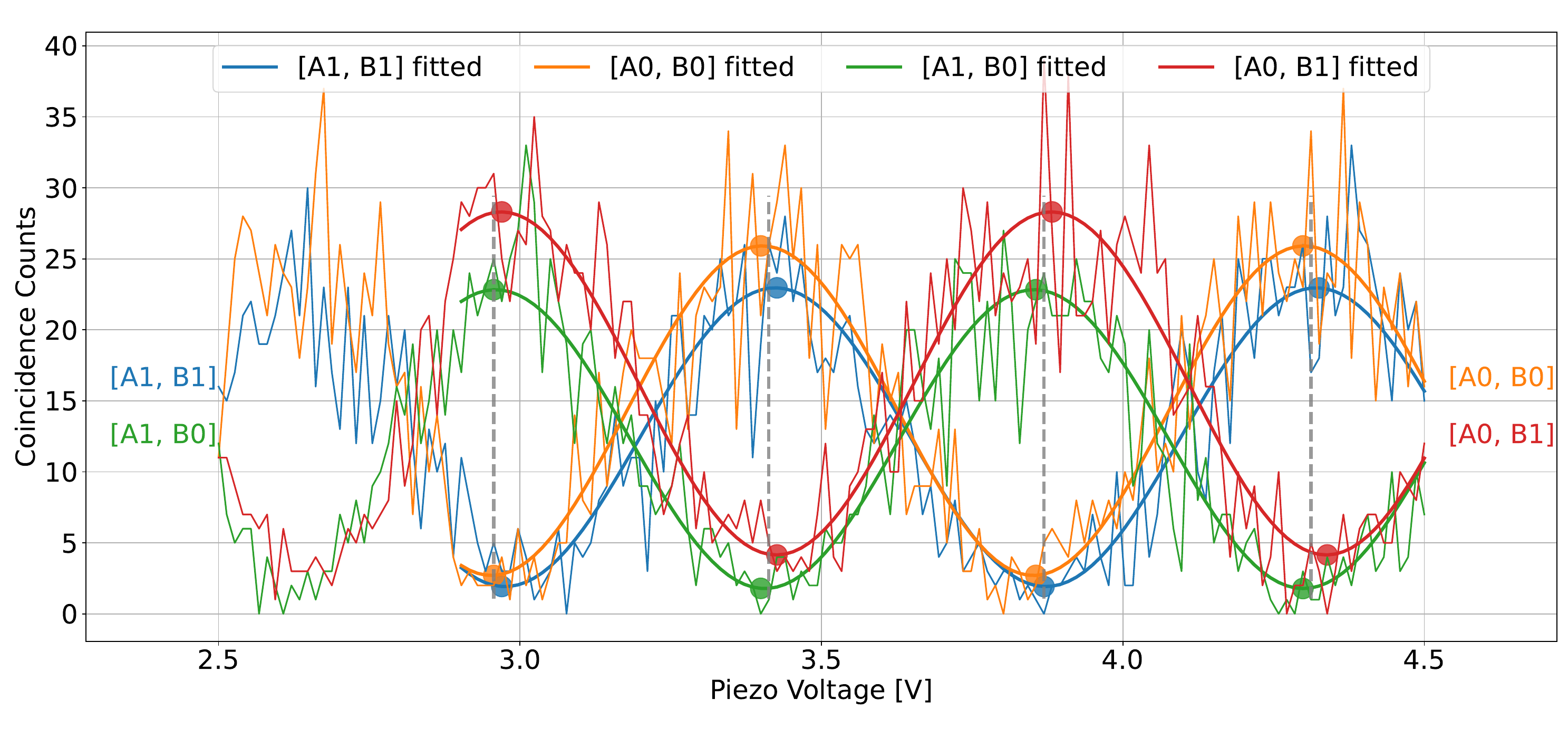}%phase_ramp_20230315-00_02_27.pdf
    \caption{Piezo voltage sweep defining the set points for phases ($\varphi_x, \varphi_y$) or,  $(\vartheta_A, \vartheta_B$). Alice and Bob set their phase references such that $\vartheta_A + \vartheta_B = 0$ for the maximum of correlated events (corresponding to the second fitted peak) which allows them to infer the XOR sum of the Referee's phase setting.}
    \label{fig:phase-scan}
\end{figure}

In the experiment, Alice and Bob use their phase shifters to set a common reference point, dependent on their classically communicated coincidences. Experimentally, this is implemented by sweeping a PZT and observing the coincidences between Alice and Bob's detectors in order to choose a voltage corresponding to one of the output correlations above. In the main text, we have assumed that Alice and Bob chose reference points, which add up to $\vartheta_A + \vartheta_B = 0$, corresponding to the second set of fitted extremal points in Fig.~\ref{fig:phase-scan}. Before each experimental run, we calibrate the Referee's PZTs in a similar fashion, noting the four voltage values pairs corresponding to 0 and $\pi$-phases, which are then applied in a random order.

% \section{Derivation of Purity and $P_{\mathrm{win}}$ with experimental imperfections}
\section{Tuning the purity}
Once the test particle (T) passes through a beamsplitter, it can be modeled as a qubit with pure state 
\begin{equation}\label{eq:qubit}
    \ket{\psi} = t\ket{1,0}_{A_T, B_T} + ir\ket{0,1}_{A_T, B_T} \equiv t\ket{0} + ir\ket{1},
\end{equation}
where $A_T, B_T$ indicate the paths towards Alice and Bob respectively and $t, r$ are the real part of the transmission and reflection coefficients of the beamsplitter. The Referee then applies his phase setting to the qubit, resulting in the state
\begin{equation}\label{eq:ref-qubit}
    \ket{\psi_R} = te^{i\varphi_x}\ket{0} + ire^{i\varphi_y}\ket{1},
\end{equation}
which is then transmitted to Alice and Bob.
In our experiment, the Referee either acts on pure quantum states or on mixed states. We model the mixed state with the density matrix
\begin{equation}
    \rho' = \int G(\varphi)\ket{\psi_R}\bra{\psi_R} d\varphi,
\end{equation}
where the mixture is the result of a Gaussian noise 
\begin{equation}
    G(\varphi) = \frac{1}{\sqrt{2\pi}\sigma} \exp(-\frac{1}{2}\left(\frac{\varphi-\varphi_x}{\sigma}\right)^2)
\end{equation}
with standard deviation $\sigma$ applied on top of the phase $\varphi_x$, set by the Referee's X PZT, to the test particle. This gives the mixed state 
\begin{equation}\label{eq:rho}
    \rho' = 
    \begin{bmatrix}
    \mathcal{T} & tr^*e^{-\sigma^2/2} \\
    t^{*}r e^{-\sigma^2/2} & \mathcal{R} 
    \end{bmatrix},
\end{equation}
which allows to compute the purity

\begin{equation}\label{eq:purs-sigma}
    \mathcal{P} = \mathrm{tr}\left[{\rho'}^2\right] = \mathcal{T}^2 + \mathcal{R}^2 + 2e^{-\sigma^2}\mathcal{T}\mathcal{R}
\end{equation}
of the mixed state, where we have defined the transmission and reflection probabilities $\mathcal{T} = |t|^2$ and $\mathcal{R} = |r|^2$. 
We can thus experimentally vary the purity of the state $\rho'$ by increasing the uncertainty, $\sigma$, on the phase setting. As a result, the purity, $\mathcal{P}$ of this state decreases from a pure state ($\mathcal{P} = 1)$.

\section{Modeling $P_{\mathrm{win}}$ in the presence of experimental imperfections}\label{sec:pwin-derivation}

Experimental imperfections can also affect the purity of the test photon input state as well as the expected win rates in the XOR game described in the main text. Imperfections can include, for example, unbalanced beamsplitters or finite distinguishability between down-converted photons. In the following, we model these imperfections to estimate our input state purity and experimental win rates.\\
Mathematically, we can express the effect of randomizing the relative phase of the input superposition state as a linear interpolation between $\ket{\psi_R}\bra{\psi_R}$ and the state with the lowest overlap that is experimentally accessible, $\ket{\psi_R^-}\bra{\psi_R^-}$, where %\footnote{since we only vary the phase $\varphi$} 

\begin{equation}\label{eq:qubit-min}
    \ket{\psi^-_R} = te^{i\varphi_x}\ket{0} - ire^{i\varphi_y}\ket{1}.
\end{equation}
This is the minimum overlap state, since we can only vary the uncertainty on the phase $\varphi_x$ to tune the purity.
We can then write the density matrices of these two states as

\begin{equation}\label{eq:rho1}
    \rho \equiv \ket{\psi_R}\bra{\psi_R} = 
    \begin{bmatrix}
    \mathcal{T} & tr^*e^{-i\varphi} \\
    t^{*}r e^{i\varphi} & \mathcal{R} 
    \end{bmatrix}
\end{equation}

\begin{equation}\label{eq:rho2}
    \rho' \equiv \ket{\psi_R^-}\bra{\psi_R^-} = 
    \begin{bmatrix}
    \mathcal{T} & -tr^*e^{-i\varphi} \\
    -t^{*}r e^{i\varphi} & \mathcal{R} 
    \end{bmatrix}.
\end{equation}
The linear interpolation between these two states then gives the mixed state density matrix

\begin{equation}\label{eq:rho'}
    \rho' = \frac{(1+\lambda)}{2}\rho + \frac{(1-\lambda)}{2}\rho^- = 
    \begin{bmatrix}
    \mathcal{T} & \lambda tr^*e^{-i\varphi} \\
    \lambda t^{*}r e^{i\varphi} & \mathcal{R} 
    \end{bmatrix}
\end{equation}
which has purity

\begin{equation}\label{eq:pur-lambda}
    \mathcal{P} = \mathrm{Tr}({\rho'}^2) = \mathcal{T}^2 + \mathcal{R}^2 + 2\lambda^2\mathcal{T}\mathcal{R}.
\end{equation}
Since the mathematical effect of randomizing the phase in state $\ket{\psi_R}$ using a Gaussian noise with standard deviation $\sigma$ is equivalent to linearly interpolating between $\rho$ and $\rho'$ with some parameter $\lambda$, we can equate the respective purities from Eq. \ref{eq:purs-sigma} and \ref{eq:pur-lambda} and find that the following must hold
\begin{equation}\label{eq:sigma-lambda}
    \lambda = e^{-\sigma^2/2}.
\end{equation}

\textit{Computing $P_{\mathrm{win}}$} ---
The probability to win the XOR game, $P_{\mathrm{win}}$, is composed of the following two scenarios: (a) where one photon is received by each party and (b) where one party receives both photons. In the latter case, the probability to guess the correct result is just 1/2, as explained in the main text, since no dependence on the relative phase exists and the parties just randomly guess the Referee's setting. Thus, to compute $P_{\mathrm{win}}$ for a specific Referee setting, we need to consider the probability for the output states from scenario (a) with the correct XOR sum (correlated or anti-correlated outputs for Alice and Bob), which we label $P^{x,y}_{\mathrm{corr}}$, and add the probability for the second scenario to occur, labeled $P_b$, multiplied by 1/2
\begin{equation}
    P^{x, y}_{\mathrm{win}} = P^{x,y}_{\mathrm{corr}} + \frac{1}{2} P_b.
\end{equation}
We can calculate the above probabilities for the mixed state $\rho'$ by taking advantage of the linearity in Eq.~\ref{eq:rho'} and calculating the individual transition probabilities for the pure states used in the decomposition of the mixture. 
More precisely, we can use the formalism presented in Ref.~\cite{aaronson_complexity_2011} where we compute the transition amplitudes for specific input ($N$) - output ($M$) state pairings by constructing the interferometer unitary, $U$, and subsequently constructing the respective submatrix $U_{MN}$, corresponding to this state pairing and computing the permanent of this submatrix.

We write our states as a vector, where each entry corresponds to the number of photons present in that mode, using the mode numbering convention laid out in Fig.~\ref{fig:mode-convention}. 
The output states where both photons arrive in the same lab are then
$$[2, 0, 0, 0],
[0, 2, 0, 0],
[0, 0, 2, 0],
[0, 0, 0, 2],
[1, 0, 0, 1],
[0, 1, 1, 0],$$
while the remaining output states correspond to one photon being detected in each lab, namely the correlated states 
$$[1, 0, 1, 0], [0, 1, 0, 1]$$
and the anti-correlated states
$$[1, 1, 0, 0], [0, 0, 1, 1].$$
Our input state is $N=[1, 0, 1, 0]$.\\
We now compute the unitary for our linear optical interferometer (including the preparation of the superposed test particle).
The unitary for a beamsplitter acting on a photon can be written as
\begin{equation}\label{eq:bs}
    U_{BS} = 
    \begin{bmatrix}
    t & ir \\
    ir & t \\
    \end{bmatrix},
\end{equation}
where $t, r$ are real coefficients. The unitary for the first beamsplitter pair with identical imbalance, acting on two, independent photons is then
\begin{equation}\label{eq:first-bs}
    U_1 = 
    \begin{bmatrix}
    t_T & ir_T & 0 & 0 \\
    ir_T & t_T & 0 & 0 \\
    0 & 0 & t_M & ir_M \\
    0 & 0 & ir_M & t_M
    \end{bmatrix}.
\end{equation}
The Referee then applies (or not) $\pi$-phases to the test photon modes (modes 1\&2), while Alice and Bob (without loss of generality) set their phases to 0, which corresponds to the matrix
\begin{equation}
    R = 
    \begin{bmatrix}
    (-1)^x & 0 & 0 & 0 \\
    0 & (-1)^y & 0 & 0 \\
    0 & 0 & 1 & 0 \\
    0 & 0 & 0 & 1
    \end{bmatrix}.
\end{equation}
Finally, the modes 1\&4 and 2\&3 are made to interfere on a pair of balanced beamsplitters 
\begin{equation}    
    U_2 = \frac{1}{\sqrt{2}}
    \begin{bmatrix}
    1 & 0 & 0 & i \\
    0 & 1 & i & 0 \\
    0 & i & 1 & 0 \\
    i & 0 & 0 & 1
    \end{bmatrix}.
\end{equation}
Taking the product of these three unitaries gives us the overall unitary for the linear optical interferometer
\begin{equation}\label{eq:nlpm-unitary}
    U = U_2RU_1 = \frac{1}{\sqrt{2}} 
    \begin{bmatrix}
    (-1)^xt_T & (-1)^xir_T & -r_M & it_M \\
    (-1)^yir_T & (-1)^yt_T & it_M & -r_M \\
    -(-1)^yr_T & (-1)^yit_T & t_M & ir_M \\
    (-1)^xit_T & -(-1)^xr_T & ir_M & t_M
    \end{bmatrix}.
\end{equation} 
Finally, we also account for the imperfect indistinguishability, which is equivalent to the HOM visibility, $\mathcal{V}$,  discussed in Sec.~\ref{sec:hom-vis}, in the case of a two-photon input and can be accounted for by linearly interpolating between the permanent and the determinant of the interferometer unitary \cite{tillmann_generalized_2015}. 
Putting this all together, we can compute the probability for a particular input ($N$) - output ($M$) state transition using the expression
\begin{align}\label{eq:transition-proba}
    P^{x, y}_{\lambda, \mathcal{V}, U, M, N} &= \frac{1+\lambda}{2} \left( \frac{1+\mathcal{V}}{2}\left|\frac{\mathrm{Per}(U^{x,y}_{M,N})}{\sqrt{m_1!...m_4!n_1!...n_4!}}\right|^2 + \frac{1 - \mathcal{V}}{2}\left|\frac{\mathrm{det}(U^{x,y}_{M,N})}{\sqrt{m_1!...m_4!n_1!...n_4!}}\right|^2 \right) \nonumber \\
    & ~+ \frac{1-\lambda}{2} \left( \frac{1+\mathcal{V}}{2}\left|\frac{\mathrm{Per}(U^{x \oplus 1,y}_{M,N})}{\sqrt{m_1!...m_4!n_1!...n_4!}}\right|^2 + \frac{1 - \mathcal{V}}{2}\left|\frac{\mathrm{det}(U^{x \oplus 1,y}_{M,N})}{\sqrt{m_1!...m_4!n_1!...n_4!}}\right|^2 \right),
\end{align}
where we have used the fact that the transformation for $\rho^-$ is the same as that for $\rho$ but replacing one of the Referee's settings with its opposite (e.g. replacing $x$ with $x \oplus 1$).

Using this, we can compute the probability, $P_b$, for both photons to arrive in the same lab.
Taking output state $M = [2, 0, 0, 0]$ as an example we can construct the submatrix
\begin{equation}\label{eq:submatrix-bunch}
    U_{MN} = \frac{1}{\sqrt{2}} 
    \begin{bmatrix}
    (-1)^xt_T & -r_M\\
    (-1)^xt_T & -r_M
    \end{bmatrix},
\end{equation}
by taking 1 copy of columns 1 and 3 for the input state $N=[1, 0, 1, 0]$ and then taking 2 copies of row 1 for the output state $M$.
Plugging this into Eq.~\ref{eq:transition-proba} yields the probability
\begin{equation}\label{eq:bunched-photon-proba}
    P^{x, y}_{\lambda, \mathcal{V}, U, M, N} = \frac{1+\mathcal{V}}{4}\mathcal{T}_T\mathcal{R}_M 
\end{equation}
for photons to bunch in one detector, where we defined the transmission and reflection probabilities $\mathcal{T}_T = |t_T|^2$ and $\mathcal{R}_M = |r_M|^2$. This probability depends solely on the indistinguishability of the two photons and the beamsplitter imbalance but not on the mixture parameter or the Referee's setting, since there is no interference effect.
Similarly, the probability for the output states $[1, 0, 0, 1]$ or $[0, 1, 1, 0]$ to occur is 
\begin{equation}\label{eq:failed-bunch-proba}
    \frac{1-\mathcal{V}}{2}\mathcal{T}_T\mathcal{R}_M \qquad \text{and} \qquad \frac{1-\mathcal{V}}{2}\mathcal{T}_M\mathcal{R}_T
\end{equation}
respectively, corresponding to the situation where the photons fail to bunch in the same lab, due to finite distinguishability.
Thus, we get the probability for both photons to arrive in the same lab
$$P_b = 2\frac{1+\mathcal{V}}{4}\mathcal{T}_T\mathcal{R}_M + 2\frac{1+\mathcal{V}}{4}\mathcal{T}_M\mathcal{R}_T + \frac{1-\mathcal{V}}{2}\mathcal{T}_T\mathcal{R}_M + \frac{1-\mathcal{V}}{2}\mathcal{T}_M\mathcal{R}_T  = \mathcal{T}_T\mathcal{R}_M + \mathcal{T}_M\mathcal{R}_T ,$$ which depends solely on the transmission and reflection probabilities of the two input beamsplitters.\\
For the scenario (a) where one photon enters each lab, let's consider the Referee setting $x=y=0$. Then the winning output states are 
\begin{equation}
    M_1 = [1, 0, 1, 0] \text{ and } M_2 = [0, 1, 0, 1].
\end{equation}
Both output states of course have the same transition probability and we find from Eq.~\ref{eq:transition-proba}
\begin{equation}
    P^{0, 0}_{\lambda, \mathcal{V}, U, M, N_1} = P^{0, 0}_{\lambda, \mathcal{V}, U, M, N_2} = \frac{1}{4} - \frac{1}{4}(1-\lambda\mathcal{V})(\mathcal{T}_T\mathcal{R}_M + \mathcal{T}_M\mathcal{R}_T).
\end{equation}
From this, we have
\begin{equation}    
    P^{0, 0}_{corr} = P^{0, 0}_{\lambda, \mathcal{V}, U, M_1, N} + P^{0, 0}_{\lambda, \mathcal{V}, U, M_2, N} = \frac{1}{2} - \frac{1}{2}(1-\lambda\mathcal{V})(\mathcal{T}_T\mathcal{R}_M + \mathcal{T}_M\mathcal{R}_T),
\end{equation}
which finally yields
\begin{equation}
    P^{0, 0}_{\mathrm{win}} = P^{0, 0}_{corr} + \frac{1}{2}P_b = \frac{1}{2} - \frac{1}{2}(1-\lambda\mathcal{V})(\mathcal{T}_T\mathcal{R}_M + \mathcal{T}_M\mathcal{R}_T)+ \frac{1}{2}(\mathcal{T}_T\mathcal{R}_M + \mathcal{T}_M\mathcal{R}_T) = \frac{1}{2} + \frac{1}{2}\lambda\mathcal{V}(\mathcal{T}_T\mathcal{R}_M + \mathcal{T}_M\mathcal{R}_T)
\end{equation}
This result is natural since the partial distinguishability is a result of tracing over distinguishable degrees of freedom (e.g. the spectrum of the photon), leading to an additional decrease of the state purity which we do not experimentally control, but which we can estimate by measuring the HOM dip visibility, $\mathcal{V}$.
Replacing $\lambda$ with its expression for the purity from Eq.~\ref{eq:pur-lambda} finally yields
\begin{equation}\label{eq:pwin-pur-complex}
    P_{\mathrm{win}} =\frac{1}{2} + \frac{1}{2}\mathcal{V}\sqrt{\mathcal{P} - (\mathcal{R}_T^2 + \mathcal{T}_T^2)} \left( \sqrt{\frac{\mathcal{T_T}}{2\mathcal{R_T}}}\mathcal{R}_M + \sqrt{\frac{\mathcal{R_T}}{2\mathcal{T_T}}}\mathcal{T}_M \right),
\end{equation}
where we have dropped the superscript, since this result holds for every Referee setting. In our experiment, the two input beamsplitters preparing the two photons in superposition have identical transmission and reflection probabilities ($\mathcal{R}_T=\mathcal{R}_M$ and $\mathcal{T}_T=\mathcal{T}_M$), leading to the simplified expression in Eq.~6, which we used to estimate the expected $P_{\mathrm{win}}$ in Fig.~2.

\section{Detection Efficiency and Normalization}

\begin{figure}
    \centering
    \includegraphics[width=1\linewidth]{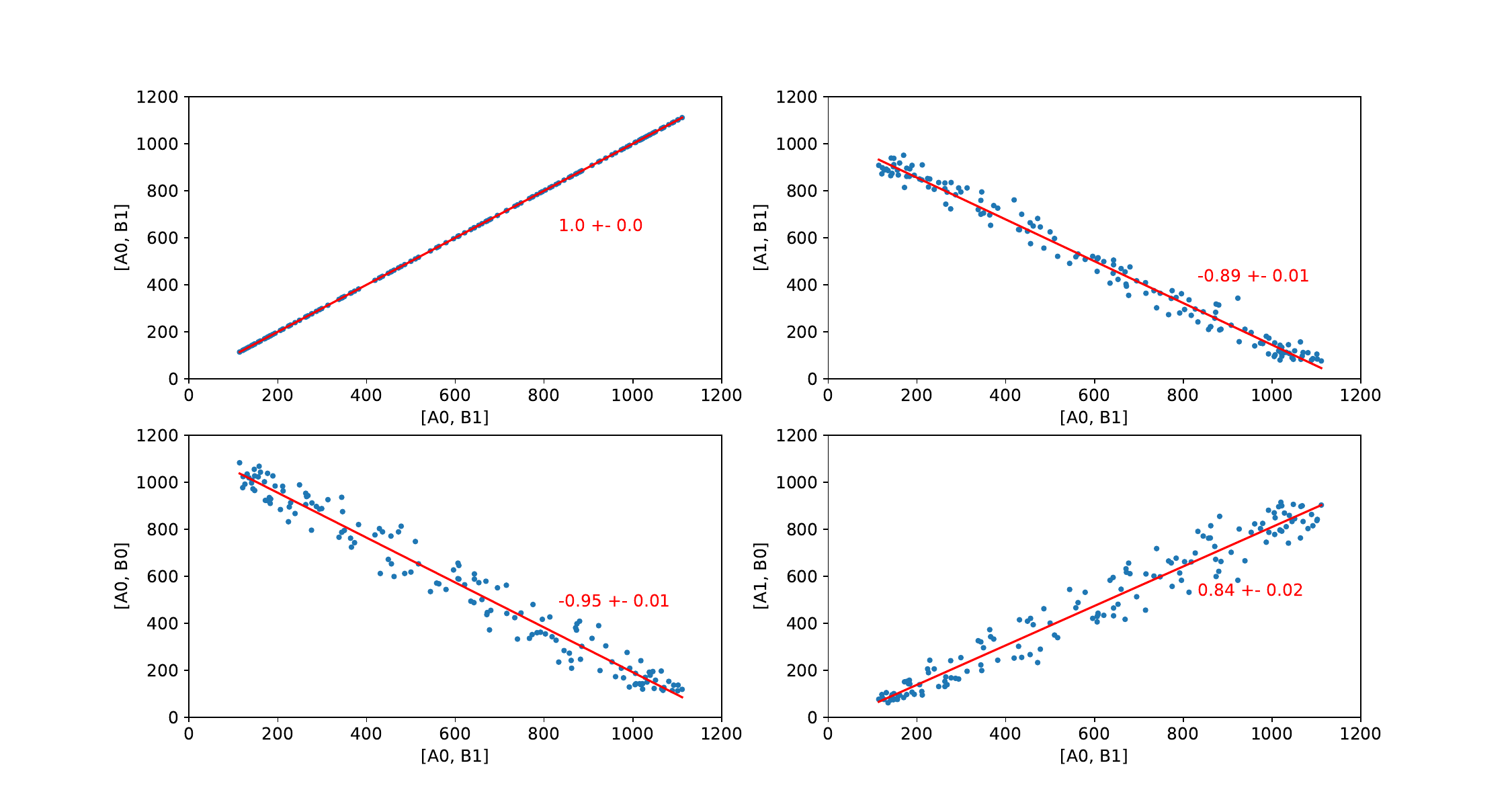}%phase_ramp_eff_20230314-22_13_39.pdf
    \caption{Relative detection efficiency is determined by performing pairwise linear regression of coincidence patterns, collected from an efficiency phase sweep with 10~s integration time.}
    \label{fig:eff-reg}
\end{figure}

To estimate the relative efficiency of each detector pair, we again sweep one of the PZTs, passing over several fringes and sampling 150 data points with an integration time of 10~s per data point. Assuming fixed total coincidence counts, we can perform pairwise linear fits between coincidence patterns (detector pairs) and use the slope relative to the most efficient coincidence pattern, $\eta_{ab}$, as the relative efficiency for each coincidence pattern, where $a,b$ are the labels for Alice and Bob's respective detectors (see Fig.~\ref{fig:eff-reg}).  

We can then compute the normalized total experimental coincidences and win coincidences,
\begin{equation}
    C_{\mathrm{tot}} = \sum_{a,b = 0}^1 \eta^{-1}_{ab}C_{ab}, \quad C_{\mathrm{win}} = \sum_{\substack{x,y=0 \\ a \oplus b = x \oplus y}}^1 \eta^{-1}_{ab}C_{ab},
\end{equation}
which finally gives the experimental win rates, normalized with the transmission and reflection probabilities and adding the winning probability from the random guessing events
\begin{equation}
    P^{(exp)}_{\mathrm{win}} = \frac{C_{\mathrm{win}}}{C_{\mathrm{tot}}}(1-2\mathcal{TR}) + \mathcal{TR}.
\end{equation}

\section{Confidence Computation}

In order to use every single detection event efficiently, we build \textit{confidence} ($\mathcal{C}$) in the nature of the resource the XOR game is played with.
Specifically, we can ask what the probability is to win the XOR game $N_\mathrm{win}$ times from $N$ games under the assumption, that we are playing with a classical resource. In statistical hypothesis testing, this represents our null hypothesis ($P_\mathrm{win} = 1/2$), which we seek to reject at a certain \textit{confidence level}. To do this, we need to compute the $p$-value for the null hypothesis.
This scenario is described by the binomial distribution 
\begin{equation}\label{eq:binom}
    P(X = N_\mathrm{win}) = \binom{N}{N_\mathrm{win}} P_\mathrm{win}^{N_\mathrm{win}} (1-P_\mathrm{win})^{N-N_\mathrm{win}}.
\end{equation}
To compute the $p$-value for winning $N_\mathrm{win}$ from $N$ games in total with $P_\mathrm{win} = 1/2$, we need to compute the cumulative distribution, $P(X \leq N_\mathrm{win})$,
\begin{equation}
\label{eq:pval}
    p\text{-value} = P( X \geq N_\mathrm{win}) = 1 - P( X < N_\mathrm{win}) = 1 - P( X \leq N_\mathrm{win}-1) = 1 - \sum_{k = 0}^{N_\mathrm{win} - 1}\binom{N}{k} \frac{1}{2^N}.
\end{equation}
\noindent We define the confidence as
\begin{equation}
    \mathcal{C} = 1 - p\text{-value} = \sum_{k = 0}^{N_\mathrm{win} - 1}\binom{N}{k} \frac{1}{2^N},
\end{equation}
which means that, if we can reject the null hypothesis with a p-value of 0.05, for example, we are 95\% confident, that our particle is not classical. Within the framework of this work, the only alternative assumption is that the XOR game is being played with a quantum resource and the confidence can thus be taken to mean, that we are 95\% confident, that the XOR game is being played with a particle in quantum superposition.

In order to correctly compute the confidence, we need to use every detection event, including the events when both photons are detected in the same lab. Since we do not use photon-number resolving detectors in this experiment, we cannot detect double-clicks. However, we can use the coincidences in Alice's $[A_0, A_1]$ and Bob's $[B_0, B_1]$ labs, to estimate the total number of unresolved double-clicks.

The probability for both photons to arrive in Alice's lab $P_{A,A}$ is given by
\begin{equation}
    P_{A,A} = P_{A_0, A_0} + P_{A_1, A_1} + P_{A_0, A_1},
\end{equation}
where a repeated index indicates an unobserved double-click and corresponds to the event from Eq. \ref{eq:bunched-photon-proba}.\\
Coincidences inside Alice's and Bob's lab are suppressed by the Hong-Ou-Mandel (HOM) effect, and will only occur due to the residual distinguishability of the photon pair with the probability from Eq. \ref{eq:failed-bunch-proba}. We can thus compute the relative occurrence of double-clicks to the corresponding number of coincidences in the lab
\begin{equation}
    \frac{P_{A_0, A_0} + P_{A_1, A_1}}{P_{A_0, A_1}} = \frac{1 + \mathcal{V}}{1 - \mathcal{V}}
\end{equation}
which allows us to compute an effective detection efficiency

\begin{align}
    \varepsilon_{2A} = \frac{P_{A_0, A_1}}{P_{A,A}} = \frac{P_{A_0, A_1}} {P_{A_0, A_1} + \frac{1 + \mathcal{V}}{1 - \mathcal{V}}P_{A_0, A_1}} = \frac{1 - \mathcal{V}}{2} = \varepsilon_{2B}
\end{align}
for two-photons-in-one-lab events. Alice (Bob) will thus only detect $\varepsilon_{2A}$ ($\varepsilon_{2B}$) of all events, that led to two clicks in their lab. For the confidence calculation, we treat the relative detection efficiency of all four Alice-Bob coincidence patterns as equal. 
We can thus account for the unobserved double-click events by discarding every Alice-Bob coincidence event with a probability of $1 - \varepsilon_{2A}$, but keeping all in-lab coincidence events.

\section{Single-Photon Source}
We use a type-II spontaneous parametric down-conversion (SPDC) source to generate photon pairs. One photon is used as the test photon and the other as the measurement resource photon, thereby ensuring a high indistinguishability. A continuous wave (CW) laser centered at a wavelength of 392~nm is set to pump a 3~mm thick beta-barium borate (BBO) crystal to generate 784~nm single photon pairs, which are filtered through a band-pass filter centered at 785~nm with a full-width half maximum of 10~nm. The crystal is pumped with a power of 120~mW and yields photon-pair rates of 15,000/s. The photons are coupled through a fiber polarization controller and pass through a half wave-plate (HWP) and a linear polarizer (LP) used for input power control before entering the nonlocal interferometer.

\section{HOM Visibility Measurement}
\label{sec:hom-vis}

We can estimate the photon indistinguishability by determining the visibility 
\begin{equation}
    \mathcal{V} = \frac{C_{\mathrm{max}} - C_{\mathrm{min}}}{C_{\mathrm{max}}}
\end{equation}
of the Hong-Ou-Mandel dip \cite{branczyk_hong-ou-mandel_2017} on Alice and Bob's detection setups respectively, by manually scanning each delay stage over a range of $300~\mu m$, thereby varying the temporal delay and collecting coincidence counts for 5~s per delay stage position.

\begin{figure}[h]
    \centering
    \includegraphics[width=.6\linewidth]{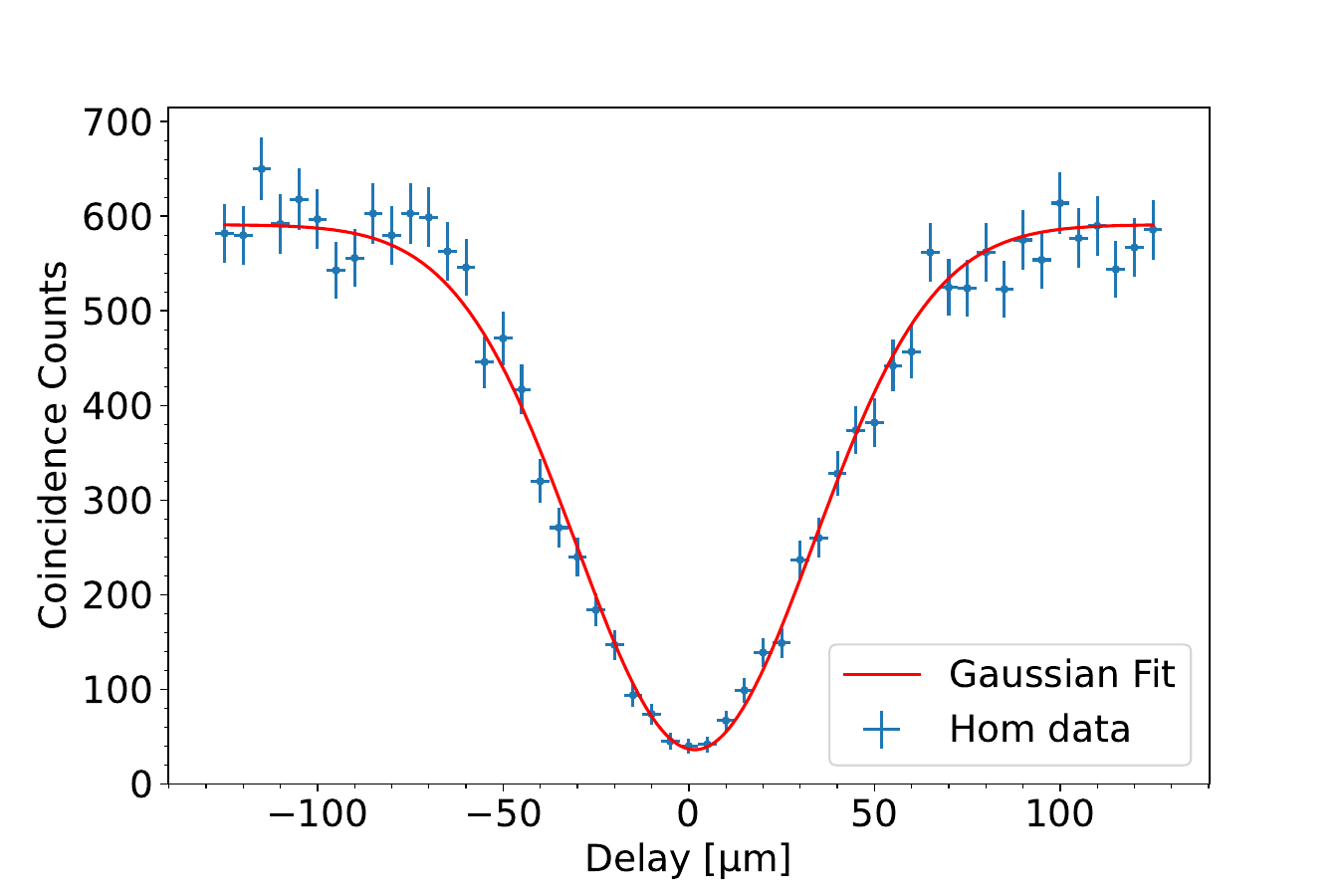}%hom_scan_20221121-15_53_32.pdf
    \caption{HOM visibility measurement on Alice's detection setup. Data points are collected with a 5~s integration time and are fitted with a Gaussian distribution.}
    \label{fig:hom-dip}
\end{figure}

We fitted the data with a Gaussian fit of the form

\begin{equation}
    G(x) = C_{\mathrm{max}} - A\exp(-\frac{(x-x_0)^2}{2\sigma^2}),
\end{equation}
where $x_0$ is the point of minimal temporal delay, $A$ is the amplitude of the Gaussian function and $\sigma$ determines the width of the fit. We find a HOM visibility of $\mathcal{V} = 94 \pm 2 \%$ on both detection setups, which we use to estimate the expected $P_{\mathrm{win}}$ in Eq.~6. The results for one of the scans can be seen in Fig.~\ref{fig:hom-dip}.

\end{document}